\documentclass[twocolumn,prX,english,nofootinbib]{revtex4}
\usepackage[utf8]{inputenc}
\usepackage{graphicx}
\usepackage{subfigure}

\def\plotwidth{8.5cm}
\def\halfplotwidth{4cm}

\providecommand{\tabularnewline}{\\}

\usepackage{babel}

\begin{document}
\title{Dynamical Heterogeneity in Lattice Glass Models}
\author{Richard K. Darst and David R. Reichman}
\affiliation{Department of Chemistry, Columbia University, 3000 Broadway, New York, NY 10027, USA.}
\author{Giulio Biroli}
\affiliation{Institut de Physique Th\'eorique, CEA, IPhT, F-91191 Gif-sur-Yvette, France and CNRS, URA 2306}
\begin{abstract}
In this paper we consider in detail the properties of dynamical
heterogeneity in lattice glass models (LGMs). LGMs are lattice models
whose dynamical rules are based on thermodynamic, as opposed to purely
kinetic, considerations.  We devise a LGM that is not prone to
crystallization and displays properties of a fragile glass-forming
liquid.  Particle motion in this model tends to be locally anisotropic
on intermediate time scales even though the rules governing the model
are isotropic.  The model demonstrates violations of the
Stokes-Einstein relation and the growth of various length scales
associated with dynamical heterogeneity.  We discuss future avenues of
research comparing the predictions of lattice glass models and
kinetically constrained models to atomistic systems.
\end{abstract}
\pacs{blah}
\maketitle

\section{Introduction}
The cause of the dramatic slowing of dynamics close to the empirically
defined glass transition is a subject of great continued interest
and debate \cite{debenedetti2001supercooled, ediger1996supercooled}.
Different theoretical proposals have been put forward
aimed at describing some or all of the phenomena commonly observed
in experiments and computer simulations \cite{garrahan2002geometrical,
garrahan2003coarse, hedges2009dynamic, lubchenko2007theory,
bouchaud2004adam, kivelson2004thermodynamic, gotze1992relaxation, 
schweizer2003entropic, dyre2006colloquium, langer2006excitation, 
heuer2008exploring}.  While these proposals are
often based on completely divergent viewpoints, many of them are able
to rationalize the same observed behaviors. This fact stems from the
somewhat limited amount of information available from experiments
and simulations. Since the growth of relaxation times in glassy systems
is precipitous, it is very difficult, and in some cases impossible,
to distinguish models solely on the basis of different predictions
of gross temperature dependent relaxation behavior. In addition, computer
simulations, which are often more detailed than experiments, are limited
by the range of times scales and sizes of systems that can be studied.
These difficulties have hindered the search for a consensus on the
microscopic underpinnings of vitrification.

Despite the continued debate that revolves around the theoretical
description of supercooled liquids and glasses, little argument exists
regarding the importance of dynamical heterogeneity as a key feature
of glassy behavior \cite{butler1991origin, hurley1995kinetic,
kob1997dynamical, yamamoto1998dynamics, schmidt1991nature,
ediger2000s}.
Dynamical heterogeneity refers to the fact that
as a liquid is supercooled, dynamics become starkly spatially heterogeneous,
requiring the cooperative motion of groups of particles for relaxation
to occur. Dynamical heterogeneous motion manifests in several ways,
and leads to violations of the Stokes-Einstein relation
\cite{chang1997heterogeneity, cicerone1996enhanced,
  jung2004excitation, xia2001diffusion, tarjus1995breakdown}, cooperative
hopping motion reflected in nearly exponential tails in particle displacement
functions \cite{berthier2005length, stariolo2006fickian,
  chaudhuri2007universal, saltzman2008large, garrahan2002geometrical},
and growing length scales such as those associated with
the recovery of Fickian diffusion \cite{swallen2009self,
  berthier2004time, pan2005heterogeneity, szamel2006time}, 
growing multi-point correlation
functions \cite{lacevic2003spatially, dasgupta1991there, franz2000non,
  berthier2007spontaneous, bouchaud2004adam, franz40analytic,
  biroli2008thermodynamic,berthier2005direct,biroli2006inhomogeneous,
  karmakar2009growing}.
Indeed, the relatively recent
explication of the phenomena of dynamical heterogeneity has dramatically
shifted the focus of the field and has placed new constraints on the
necessary ingredients for a successful theory of glass formation.

Given the similarity of some aspects of dynamical heterogeneity to
critical fluctuations in standard critical phenomena, it is natural
to investigate two and three dimensional simplified coarse-grained
models that encode the crucial
features of this heterogeneity. Currently, the most investigated class
of coarse-grained models are the ``kinetically constrained models''
(KCMs) \cite{garrahan2002geometrical, garrahan2003coarse,
  fredrickson1984kinetic, kob1993kinetic, ritort2003glassy}. 
KCMs are spin or lattice models that generate slow, glassy
relaxation via constraints on the dynamical moves that are allowed.
The slowing down of the dynamics is caused by rarefactions of
facilitating regions, also called defects.
Importantly, although the dynamics is complex the thermodynamics is
trivial since
the dynamical rules are such that all configurations are equally likely.
The philosophy of this viewpoint is that thermodynamic quantities,
such as the configurational entropy, are not the fundamental underlying
cause of the growing time scales in supercooled liquids.
It has been argued that the quantitative disagreement \cite{biroli2005defect}
between
thermodynamic features of KCMs and real experiments is of little
dynamical consequence \cite{chandler2005thermodynamics}.
In support of this perspective is the
fact that KCMs have been remarkably successful in generating features
of dynamical heterogeneity such as Stokes-Einstein decoupling, growing
dynamical length scales, and excess tails in the real-space particle
displacement function \cite{jung2004excitation, pan2005heterogeneity,
   hedges2008dynamic, hedges2007dynamic}.

\begin{figure}
\centering
\includegraphics[width=\plotwidth]{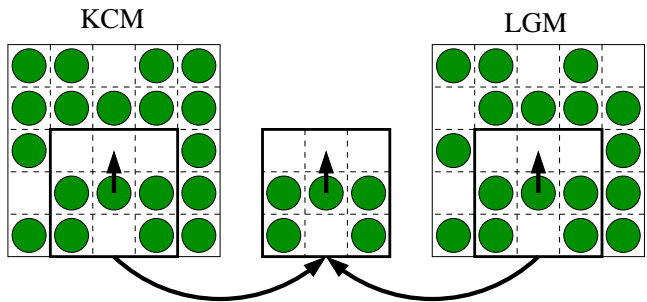}
\caption{Comparison and distinction of a caricature of a kinetically
  constrained model with a lattice glass model.  In the KCM any
  configuration is allowed, but move may only be made if a particle
  has at least one missing neighbor before and after the move.  In the
  LGM, the global configuration is defined such that all particles
  must have at least one missing neighbor, and all dynamical
  moves must respect this
  rule.  Note that the local environment around the moving particle is
  identical in this example, while the global configurations are
  distinct.  Periodic boundary conditions are assumed for both panels.}
\label{fig:modeldemo}
\end{figure}

On the other hand, one may wonder if a deeper viewpoint would allow
for an understanding of the kinetic rules that govern particle motion
in the supercooled liquids.
It is natural to speculate that such aspects might have roots
in the thermodynamics of configurations. Indeed, simple local Monte
Carlo ``dynamics'' can reproduce all features of dynamical heterogeneity
seen in Newtonian molecular dynamics simulations, and are based simply
on making local moves that are \emph{configurationally} allowed
\cite{berthier2007revisiting}. Lattice
models based on this concept are called ``lattice glass models''
(LGMs), and were first considered by Biroli and M\'ezard
\cite{biroli2001lattice}. The rules
for such models seem at first sight like that of KCMs.  For example
in the simplest versions of such models a particle may move if it
is surrounded by no more than a fixed number of nearest neighbors
before and after the move \cite{mccullagh2005finite, dawson2003dynamical,
ciamarra2003lattice,  pica2003monodisperse}.
Locally this is identical to the type of
dynamical constraint that appears in the KCMs introduced by Kob and
Andersen \cite{kob1993kinetic}.
However, this
constraint must be met globally: \emph{all} particles must have
no more than a fixed number of nearest neighbors. As the density of the system
increases, fewer and fewer configurations exist for which these constraints
may be satisfied. It is thus the entropy of configurations that governs
the slowing of dynamics, intimately connecting the non-trivial thermodynamic
weight of states accessible to the local dynamics.
Indeed, LGMs can be solved exactly within the Bethe approximation, or
on Bethe lattices \cite{biroli2001lattice, ciamarra2003lattice},
and have been shown to have a glass transition due
to the vanishing of the configurational entropy.
The distinction between the KCM and LGM viewpoint is illustrated in
Fig. \ref{fig:modeldemo}.

LGMs have been studied by a number of groups, but the focus has not
generally been on real-space aspects of dynamical heterogeneity. For
example, Coniglio and coworkers have developed a simple LGM that avoids
crystallization and displays many features of typical glass-forming
materials, including a growing multi-point susceptibility
($\chi_{4}(t)$) \cite{ciamarra2003lattice, pica2003monodisperse}.
On the other hand, this system appears to behave as a strong glass-former,
with a stretching parameter close to one, and exhibits essentially
no Stokes-Einstein violation. Our goal in this work is to survey in
detail the dynamical behavior of a new LGM which bears similarity
to the original Biroli-M\'ezard model but is not prone to crystallization.
The main conclusion that we draw is that LGMs are at least as realistic
as KCMs in their description of all commonly studied features of dynamical
heterogeneity. In this regard, simple coarse-grained lattice models
based on the thermodynamic weight of states are no less viable as
fundamental caricatures of glassy liquids than are KCMs based
on weights of trajectories. We conclude our work by highlighting several
key ways that LGMs and KCMs may be distinguished. We reserve the investigation
of these comparisons for a future study. Our paper is organized as
follows:  Sec.~\ref{sec:model} outlines the model.  Sec.~\ref{sec:dynamics}
discusses both simple averaged dynamics as well as
aspects of dynamical heterogeneity.  In Sec.~\ref{sec:conclusion} we
conclude with a discussion of the meaning of our findings and the
future directions to be pursued.

\section{Model}
\label{sec:model}
Here we define the LGM that forms the basis of our simulations. The
original model of Biroli-M\'ezard is quite prone to crystallization
\cite{biroli2001lattice}.
This fact makes its use problematic for the study of glassy behavior
since crystallization always intervenes before supercooling becomes
significant. The crystallization problem persists on a square lattice
for all binary mixtures we have studied. However, we have found that
certain generalizations of the Biroli-M\'ezard model with three species
of particles are stable against crystallization for the densities that
are sufficiently high that glassy dynamics may be clearly observed.

Our model follows the original rules of the Biroli-M\'ezard model. Particles
exist on a cubic periodic lattice of side $L=15$ and each lattice
site can contain
only zero or one particle. All particles, at all times, must satisfy
the condition {\em a particle of type ``$m$'' must have $m$ or fewer
neighbors of any type.} A neighbor is considered
any particle in one of the $2d$ ($d$=dimensionality) closest lattice sites
along the cubic coordinate axes
\footnote{It should be noted that LGMs of the type described here
  involve extreme constraints that must be globally satisfied and are
  thus not realistic translations of off-lattice particle-based
  models.  Such constraints might indeed induce artificial behavior,
  especially at higher densities. It would be most interesting to
  investigate ``soft'' versions of such models where constraints may
  be locally violated at the cost of an energy penalty. In this
  regard, such models would be the configurational analog of KCMs
  where dynamical constraints may be broken at the cost of an energy
  penalty, see Chandler, D. and J.P. Garrahan, ``Dynamics on the Way
  to Forming Glass: Bubbles in Space-time'' arXiv:0908.0418v1;
  Submitted to Annual Reviews of Physical Chemistry (2010).}.

The particular three species model we employ is defined by 10\% type
1 particles, 50\% type 2 particles, and 40\% type 3 particles. We
denote this model the ``t154'' model to indicate its basis in
thermodynamics and to specify the types and percentages of each particle.
The composition of t154 model was determined via trial and error by
picking particle types with clashing crystallization motifs thereby
frustrating crystallization.
Crystallization was monitored by inspection
of the angle resolved static structure factor, direct inspection of
configurations, and by monitoring bulk thermodynamic quantities.

As discussed in the introduction, there appear to be strong
similarities between the rules that govern
KCMs such as the Kob-Andersen model and the t154 model
\cite{kob1993kinetic}. For example both models employ constraints with a
maximum number of neighbors, but in the Kob-Andersen model this restriction
only applies to the \textit{mobile} particles, while in the t154 model
applies to \textit{all} particles. Our model does not require any
special dynamics methods. We employ local canonical Monte Carlo {}``dynamics''
via primitive translational moves \cite{berthier2007revisiting}.
Note that for the t154 model the
energy can only be zero (no packing violations) or infinite (packing
violation or overlap), thus the acceptance criteria reduces to rejection
if there is a packing violation and acceptance otherwise. This allows
us to implement an event-driven algorithm which accelerates the simulation
of lattice dynamics \cite{bortz1975nam}.

\begin{figure}
\centering
\subfigure{
  \includegraphics[angle=-90,width=\plotwidth]{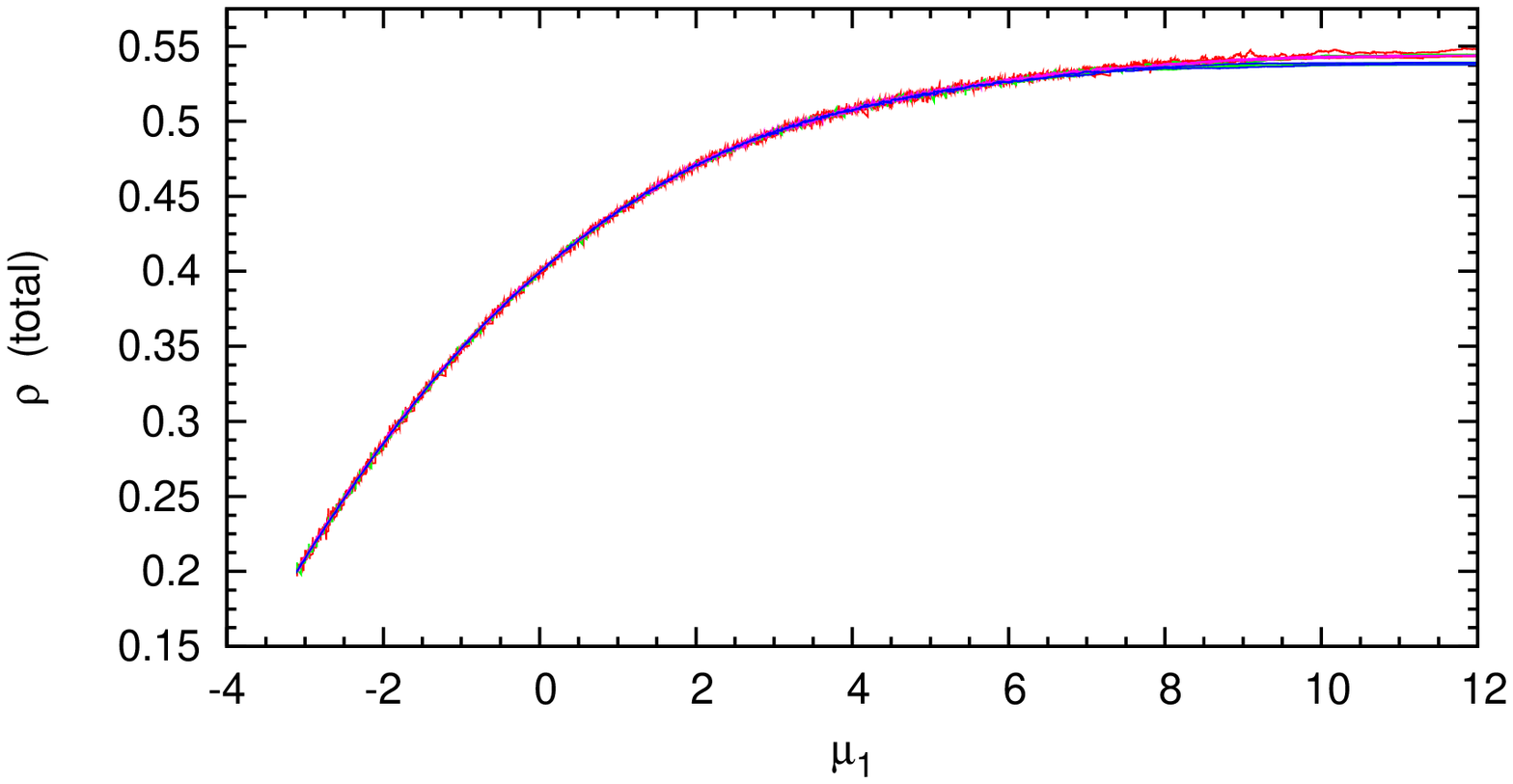}}
\subfigure{
  \includegraphics[angle=-90,width=\plotwidth]{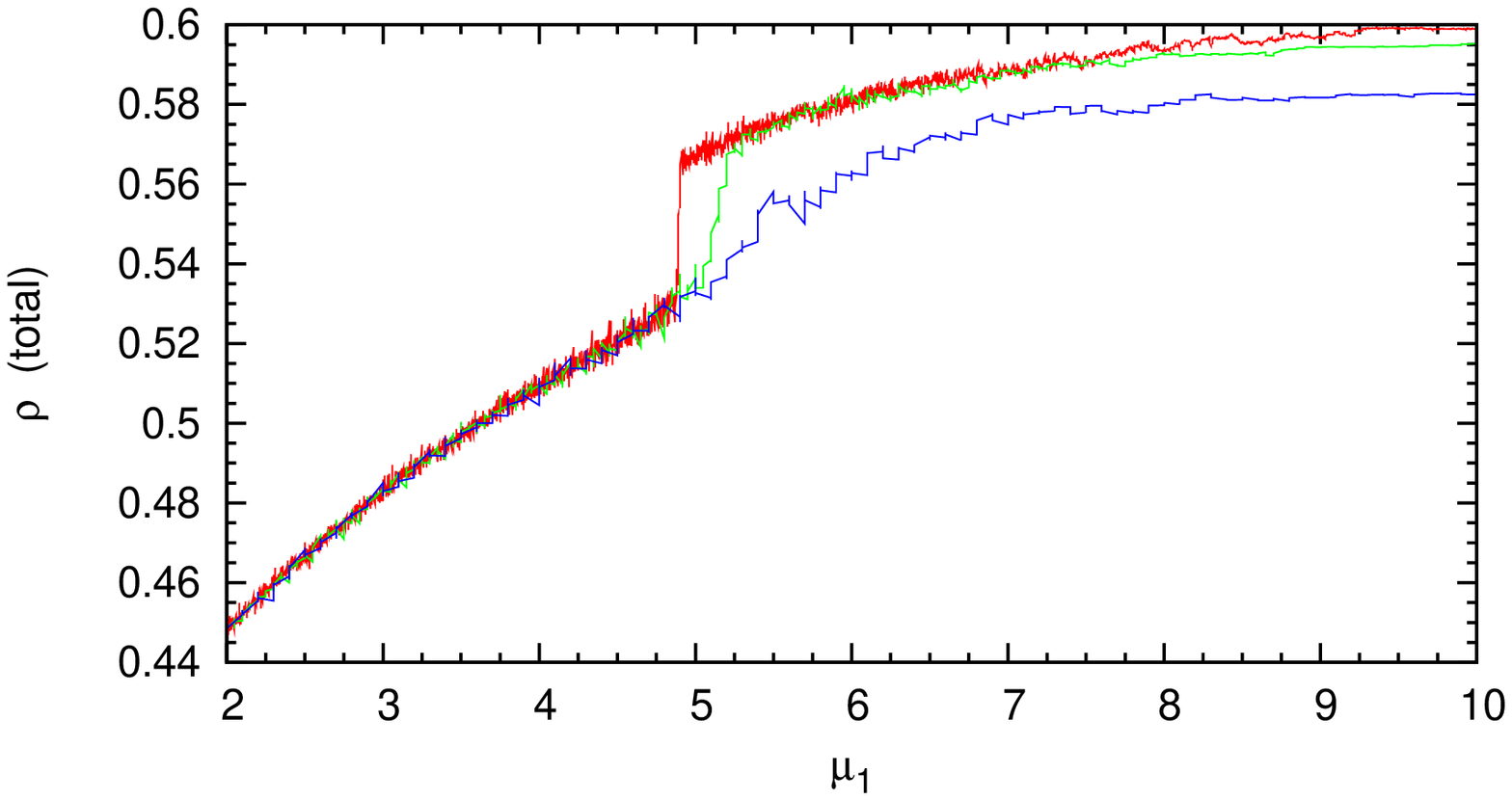}}
\caption{Crystallization thermodynamics in LGM.
\textbf{Top:} The t154 model.  $\mu_1$ refers to the chemical
potential of the type 1 particles.  The
maximum density observed for the 15$^{3}$ lattice is .5479 (exactly
1849 out of 3375 lattice sites occupied). The three plotted quenching
rates vary between a .01 and .05 increase of $\mu_1$
per 10000 cycles. \textbf{Bottom:} A close up of the equivalent plot
for the BM model. Note the clear discontinuity upon crystallization.
Slower $\mu$-increase rates produce a sharper discontinuity.}

\label{fig:crystcurve}
\end{figure}

For thermodynamic studies we employ grand-canonical Monte Carlo with
both translational moves as well as particle insertion/deletion. Fig.
\ref{fig:crystcurve} contains a plot of the density of the system
as a function of the chemical potential of type 1 particles. Models
which crystallize (such as original binary model of Biroli-M\'ezard)
have a sharp jump in this curve at the crystallization point. Clearly,
this feature is absent in the t154 model. For comparison, both curves
are displayed
\footnote{A subtle issue arises in the nature of glassy behavior
  observed in the t154 model outlined in this work. LGMs could have a
  dynamical percolation-like transition, as in the spiral model
  \cite{biroli2008spiral}. This has been indeed found in some LGMs on
  the Bethe lattice \cite{rivoire2003glass} and would slow down the dynamics
  for reasons completely different from the diminishing of the
  configurational entropy.  If there is a low-lying crystal phase then
  one can show that this dynamical percolation-like transition cannot
  take place in finite dimension.  Although we have not found a
  crystal phase for the model, the existence of such a transition
  seems unlikely and irrelevant for our present work. First, it can be
  shown that blocked structures, if they exist, have to verify much
  more constraints than in the spiral model \cite{biroli2008spiral}.
  Second, we have found that the relaxation time growth of the
  persistence functions with increasing density in local canonical
  Monte Carlo simulations are similar to those under grand-canonical
  dynamics, which cannot contain any blocked structure.  The union of
  these two facts render the dynamical blocking scenario highly
  unlikely.}.

\section{Dynamical Behavior}
\label{sec:dynamics}

\subsection{Simple Bulk Dynamics}

\begin{figure}
\centering
\subfigure{
  \includegraphics[angle=-90,width=\plotwidth]{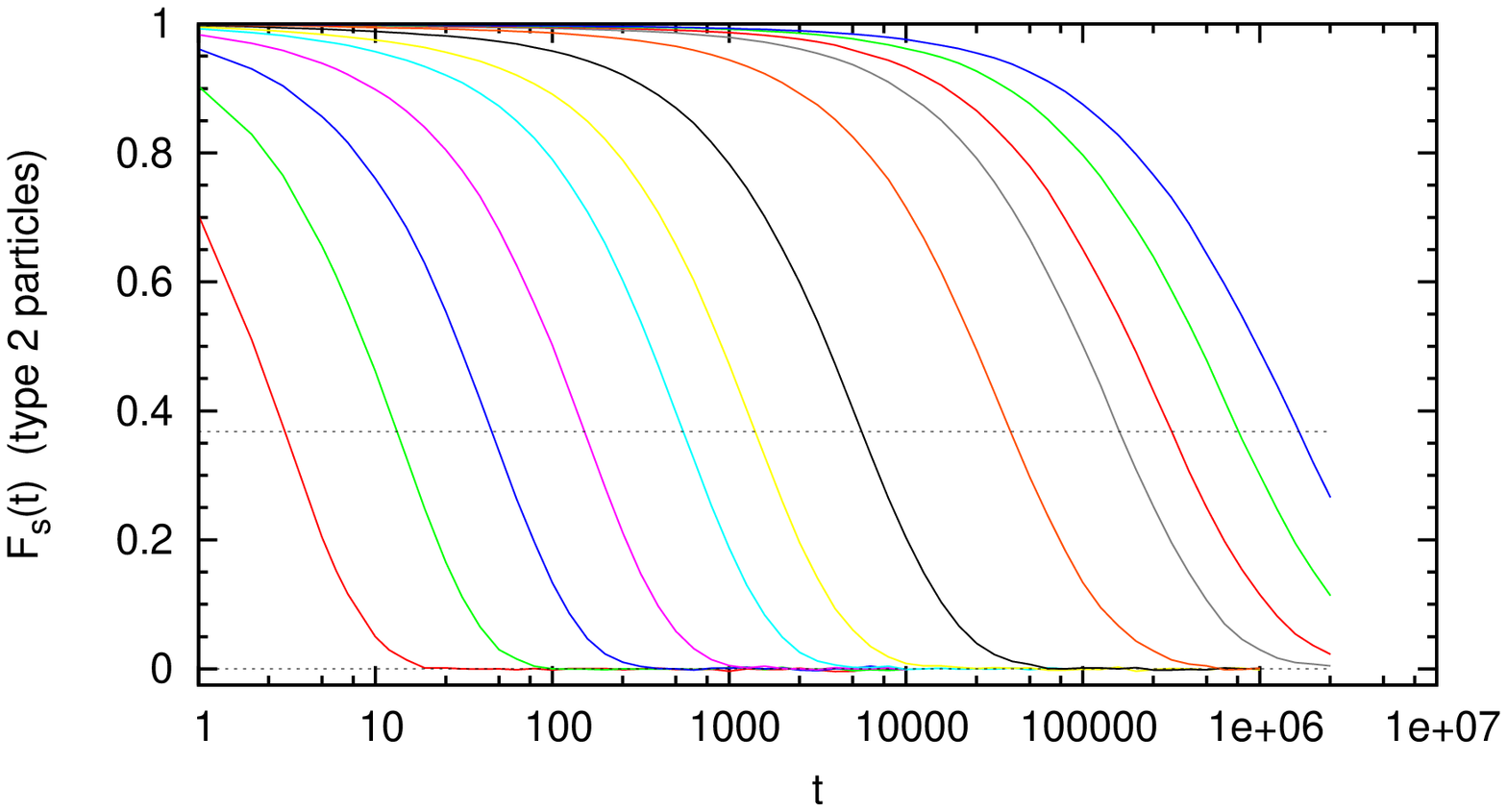}}
\subfigure{
  \includegraphics[angle=-90,width=\plotwidth]{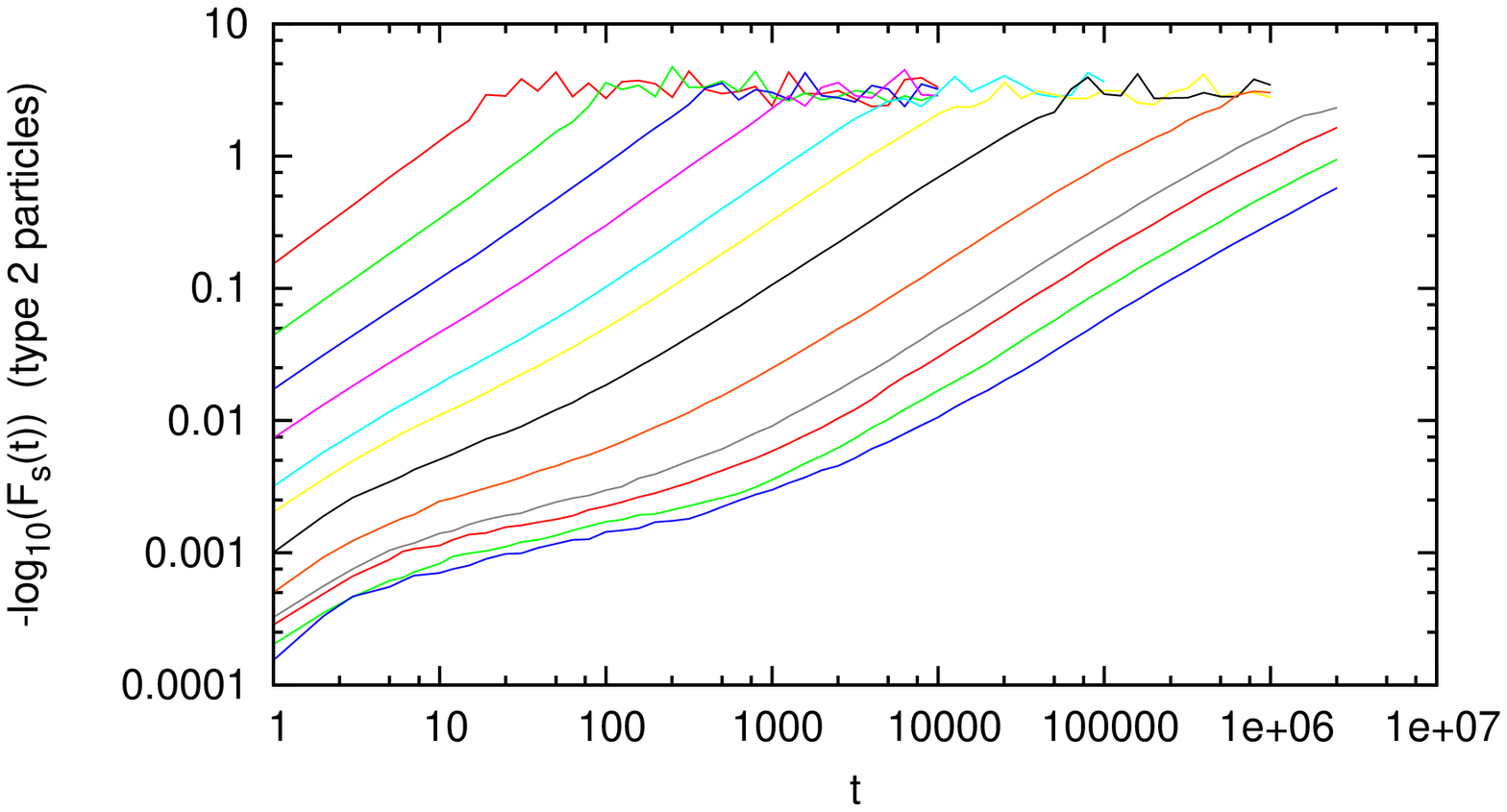}}
\caption{Decay of the self-intermediate scattering function $F_{s}(k,t)$
for $k'=5$ ($k = \frac{2\pi}{L}k'$). Densities are .3,
.4, .45, .48, .50, .51, .52, .53, .535, .5375, .5400, .5425 from fastest
relaxation to slowest relaxation. These densities are used in all
plots in this paper unless otherwise indicated. \textbf{Top:} Plotted
on a linear-log scale. \textbf{Bottom:} Same data as upper panel
plotted on a $\log(-\log_{10}(F_{s}(k,t)))$
vs $\log(t)$ scale. Lowest density curves are at the
top left.}
\label{fig:Fs_decay}
\end{figure}

In this subsection we describe the behavior of a simple 2-point
observable, namely the self-intermediate scattering function
\cite{balucani1994dynamics}, defined
as \begin{equation}
  F_{s}(k,t)=\left<\frac{1}{N}\sum_{i}e^{i\mathbf{k}
      \cdot\left[\mathbf{r}_{i}(t)-\mathbf{r}_{i}(0)\right]}\right>.
  \label{eq:Fs_self}
\end{equation}
We measure $F_{s}(k,t)$ only for the type-2 particles which are
present in the greatest fraction for the three distinct species.
Throughout this paper, we report $k$-vectors using $k'$, where $k =
\frac{2\pi}{L}k'$. We
have checked that $F_{s}(k,t)$ is qualitatively similar for the other
species of particles. The relaxation of $F_{s}(k,t)$ of the system at
the wavevector $k'=5$ ($k=\frac{2\pi}{3}$) for various densities is shown in
Fig. \ref{fig:Fs_decay}. The bulk of the decay may be fit to a
stretched exponential function,
$F_{s}(k,t)=\exp(-(t/\tau_{\alpha}(k))^{\beta(k)})$.  As is customary,
the alpha-relaxation time is found by the value
$F_{s}(\tau_{\alpha})=1/e$ and the $\beta(k)$ exponent is determined by a
direct fit to the terminal decay. We find that for densities below
approximately $\rho=0.48$ the value of $\beta$ saturates at the
expected value $\beta=1$ characteristic of simple non-glassy dynamics,
while for the highest density simulated, $\beta=0.7$.  This behavior,
over a similar range of supercooling, is reminiscent of the behavior
found in atomistic models of glass-forming liquids
\cite{kob1995testing, wahnstrom1991molecular}.  In order to
better reveal the relaxation behavior, $F_{s}(t)$ is also displayed on
a log-log vs.\ log-time scale. In this plot, the slope of the long time
growth is related to the exponent $\beta.$ We have found that the
values of $\beta$ extracted from the slopes of the long time portion
of the log-log vs.\ log plot indeed coincide with that found by a
direct fit to a stretched exponential form. At the highest densities a
shoulder appears in the short time relaxation. This feature is
indicative of a secondary relaxation feature perhaps akin to
beta-relaxation in realistic glass-forming liquids.  It should be
noted, however, that the amplitude of this feature is very close to
unity. This is quantitatively distinct from the plateau values
expected in atomistic off-lattice models \cite{kob1995testing,
  wahnstrom1991molecular} and even LGMs with more
complicated lattice degrees of freedom \cite{ciamarra2003lattice,
  pica2003monodisperse}, but is similar to that
encountered in simple spin models such as variants of the Random
Orthogonal Model \cite{sarlat2009predictive}.

\begin{figure}
\centering
\subfigure{
  \includegraphics[angle=-90,width=\plotwidth]{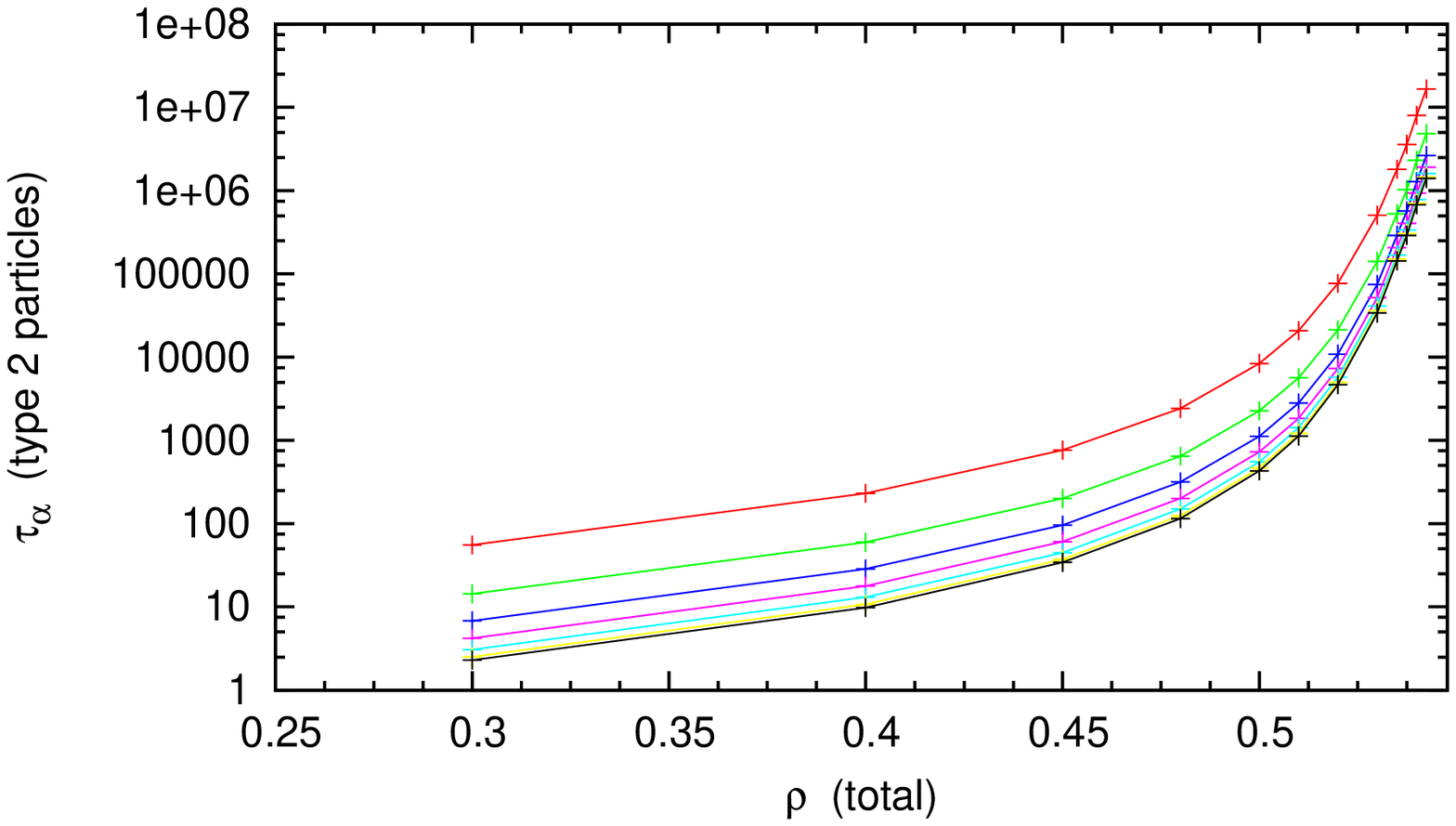}}
\subfigure{
  \includegraphics[angle=-90,width=\plotwidth]{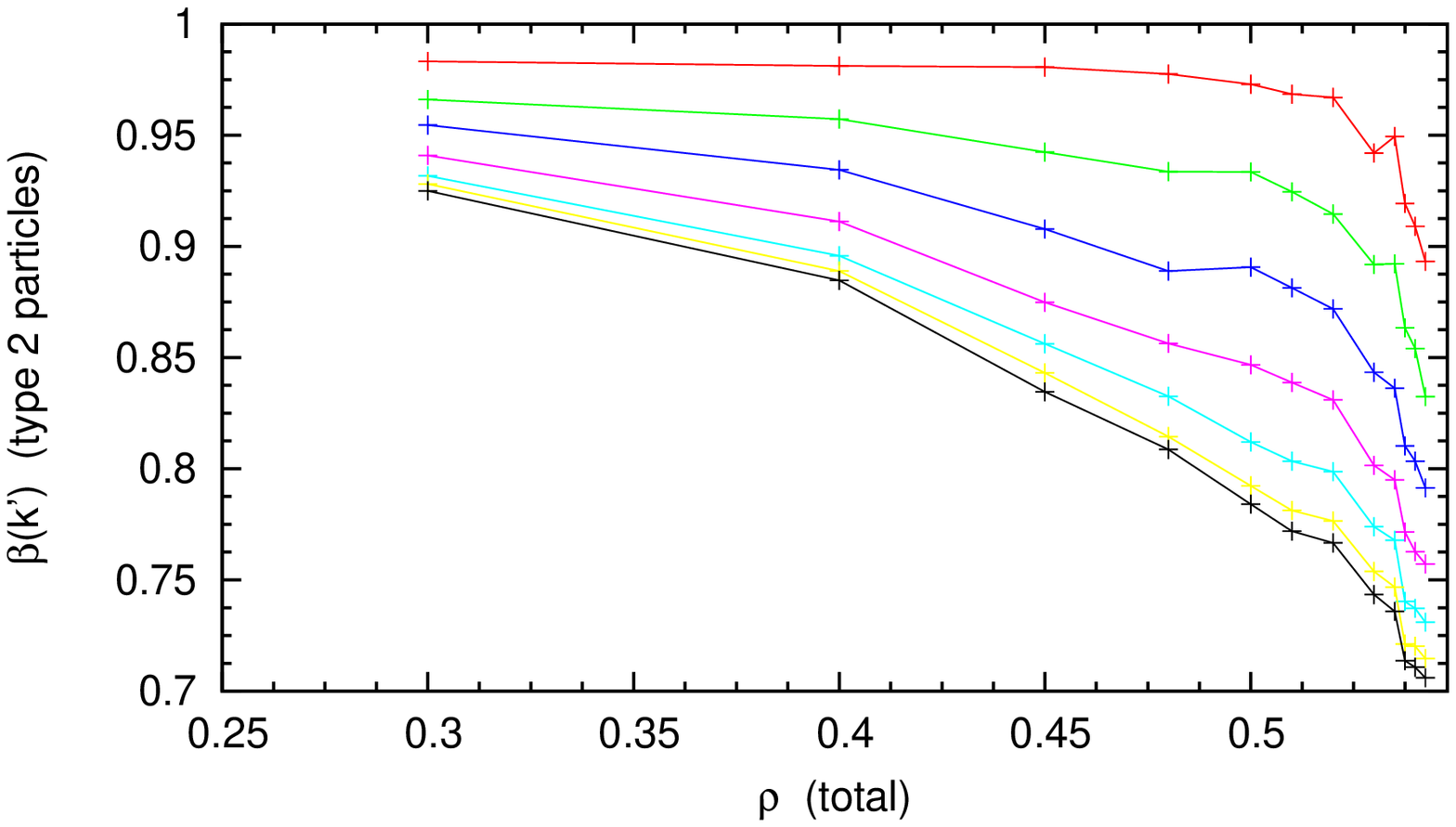}}
\subfigure{
  \includegraphics[angle=-90,width=\plotwidth]{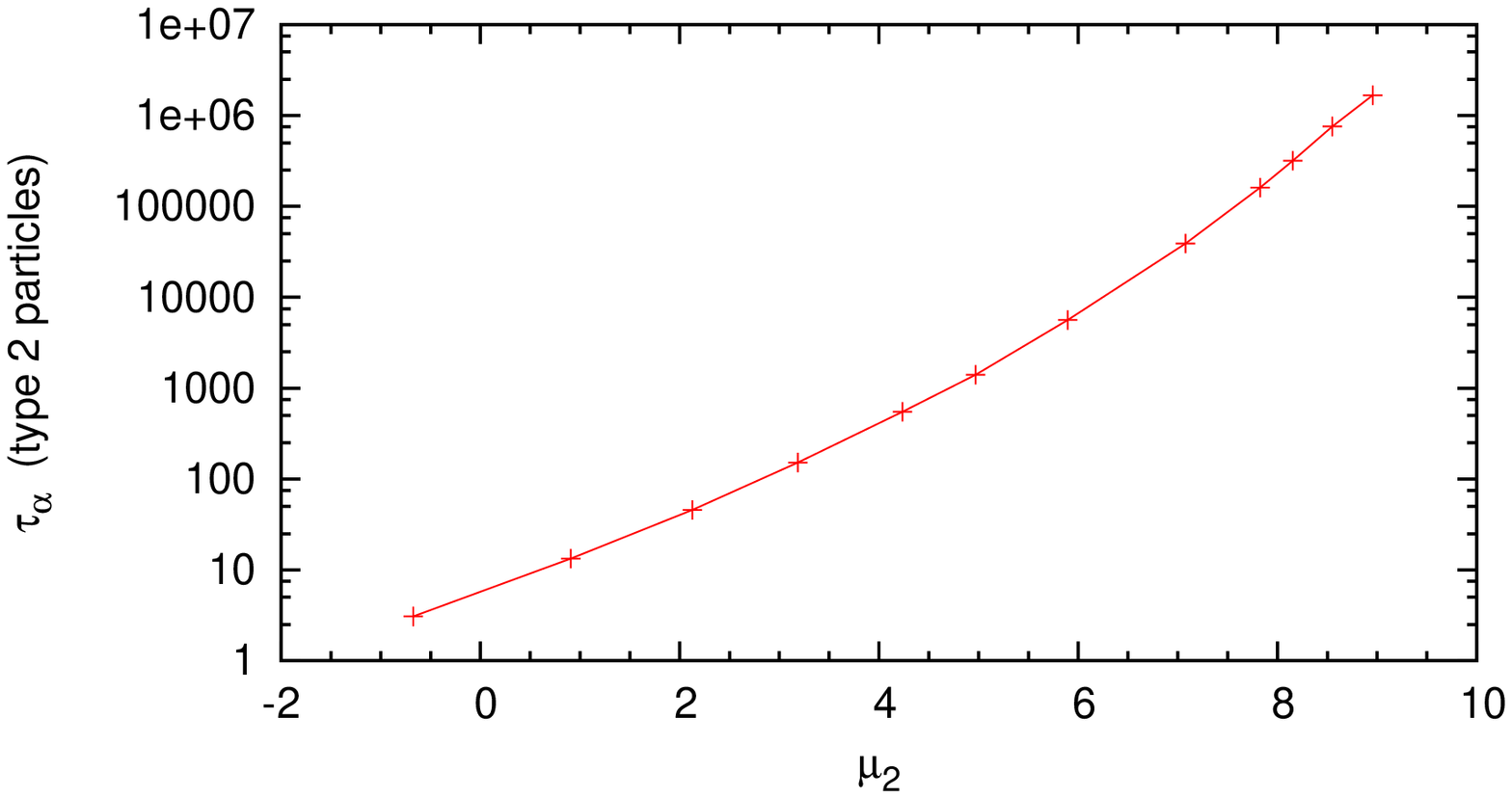}}
\caption{\textbf{Top:} $\tau_\alpha$ (time at which $F_{s}(k,t)=1/e$)
as a function of density, $\rho$. Plotted for $k'=1,2,3,4,5,6,7$,
with lowest $k$ at the top.
\textbf{Center:}
Beta stretching exponent of $F_{s}(k,t)$
(from terminal fits $F_{s}(k,t)\sim\exp(-(t/\tau_{\alpha})^{\beta})$).
Lowest $k$ curve is at the top of the plot.
\textbf{Bottom:} Plot of log scale $\tau_\alpha$ against chemical
potential $\mu$ of type 2 particles.  The behavior is consistent with
$\tau_\alpha=5.7\exp(-21\mu_2/(\mu_2-24))$.}
\label{fig:tau}
\end{figure}

As is typical of fragile glass-forming systems, the t154 model exhibits
relaxation times that do not follow the (generalized) Arrhenius form
\cite{berthier2009compressing}.
This behavior is illustrated in Fig. \ref{fig:tau}. At low densities,
plots of $\log(\tau)$ versus $\rho$
indeed follow a straight line, however in the vicinity of $\rho
\sim0.5$ the plot of $\tau$ versus
$\rho$ deviates from this straight line and the
functional density dependence of the relaxation time becomes much
more precipitous. While we have not attempted to quantitatively characterize
this density dependence, it should be noted that the onset of increased
sensitivity to changes in density occurs is the same narrow window
that marks the noticeable decrease in the values of the stretching
exponent $\beta$.


\subsection{Motion on the Atomic Scale}

We begin our discussion of the nature of heterogeneous dynamical behavior
in the t154 LGM by observing the qualitative details of particle motion
under supercooled conditions. This will set the stage for analysis
of quantitative measures of dynamical heterogeneity in the model. For
the sake of comparison, we also investigate the analogous behavior
in the Kob-Andersen model. This comparison is useful because it suggests
how models with similar local rules but different global rules (rooted
in either the purely kinetic or thermodynamic basis of the particular
model) may give rise to distinct dynamics at the particle scale.

\begin{figure}
\centering
\subfigure[]{
  \includegraphics[width=\halfplotwidth,angle=-90]{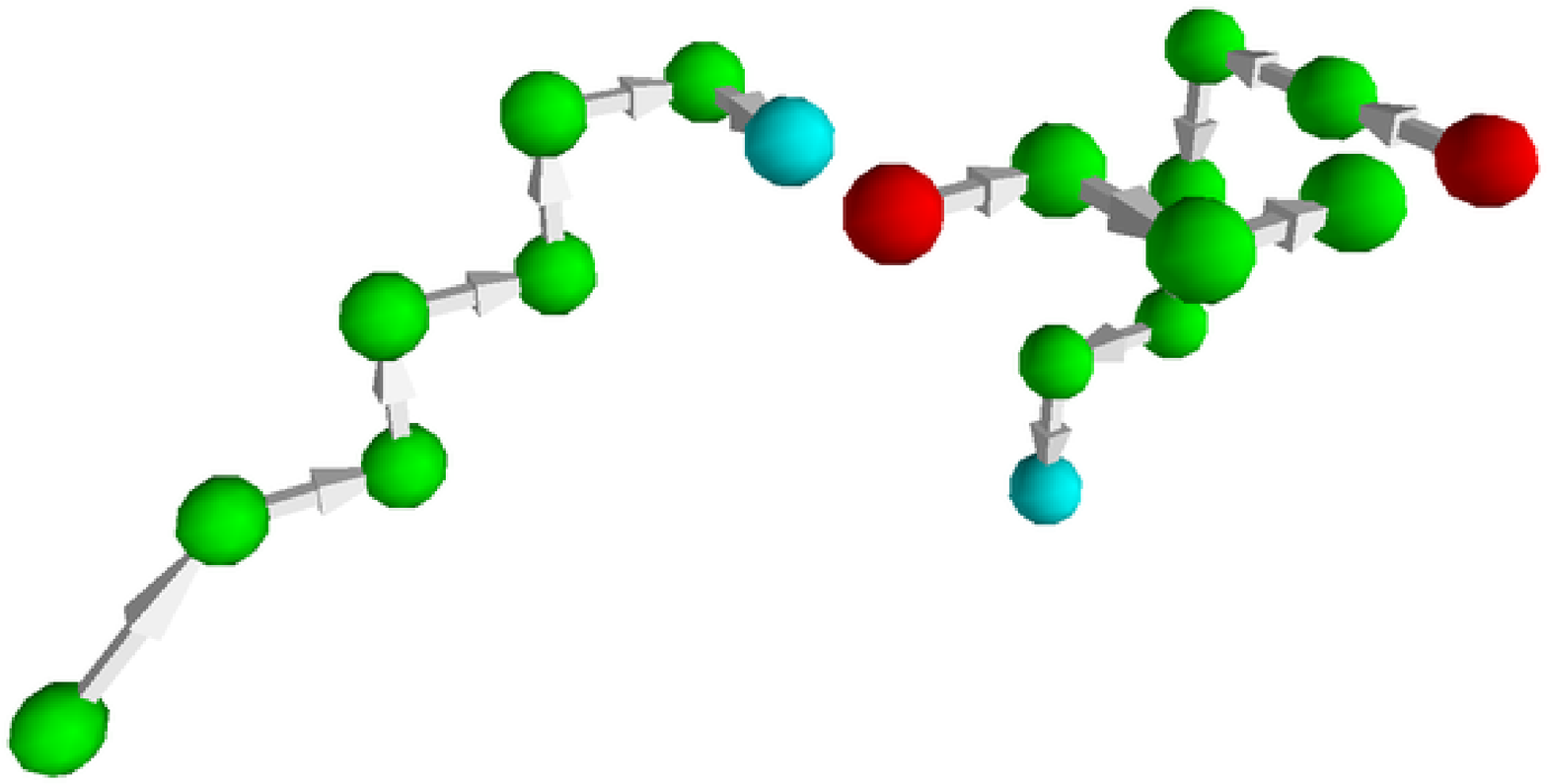}}
\subfigure[]{
  \includegraphics[width=\halfplotwidth,angle=-90]{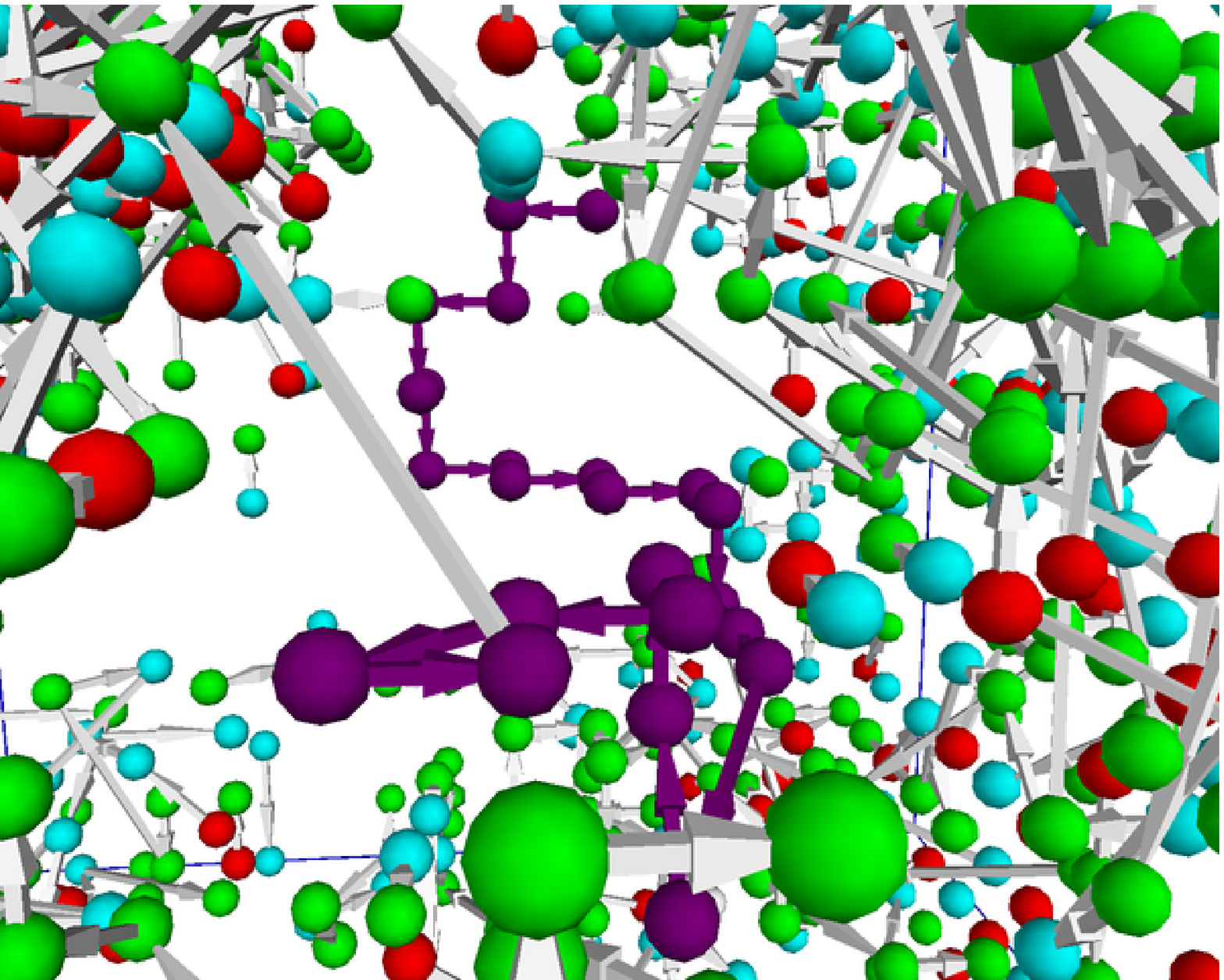}}
\caption{Examples of string-like motion apparent in the t154 model.
\textbf{(a)} An example
of a string with all neighboring particles removed.
\textbf{(b)}
A similar string in the context of other particles. Note that here the string
is truly isolated in space, away from other mobile particles. In these
figures, type 1 particles are white, type 2 particles are blue, and
type 3 are green. Sites occupied at the initial time but vacated at
the final time are shown in red. These pictures show only the differences
in position of particles between the origin of time and the final time, not
the path the particles took to achieve that displacement. All figures
are at a density of .5400, with $\Delta t$ times  in (a) 251,
(b) 199526. The $\alpha$-relaxation time for $k'=5$ at this density
is about $7.8\times10^{6}$}
\label{fig:strings}
\end{figure}

We start by simply observing the patterns of mobility in real space
starting from a set initial condition of the t154 model found at a
given density after equilibration. A similar analysis has been performed
recently by Chaudhuri {\em et al.} for the Kob-Andersen
model, where no equilibration is required since all initial configurations
with a set density of defects are allowed
\cite{chaudhuri2008tracking}. For a theoretical description of the
dynamics of the Kob-Andersen
model, see \cite{toninelli2004spatial}. We note that, as expected,
the t154 model exhibits regions of spatially localized particle activity
against a backdrop of transiently immobilized particles. A rather
remarkable feature of the patterns of mobility in this model is that
we find evidence of string-like motion, where a group of particles
moves over a short distance, each taking the place of the previous
particle in the string \cite{donati1998stringlike,
  donati1999spatial}. This motif can be seen mostly on timescales
less than the $\alpha$-relaxation time, but occasionally string-like
motion may be seen to persist on longer timescales. This behavior
is demonstrated in Fig. \ref{fig:strings}.

The behavior of particle motion observed in the Kob-Andersen model
is somewhat different than that seen in the t154 model as described
above. As in the t154 model, and as observed by Chaudhuri {\em
et al.}, motion in the Kob-Andersen model shows similar activity
regions in the vicinity of defect sites giving rise to heterogeneous
motion. However, the boundaries between active and inactive regions
at comparable timescales appear to be more distinct in the Kob-Andersen
model. Furthermore, the particle scale motion in the Kob-Andersen
is much more isotropic, exhibiting much fewer cases of directional
mobility compared with the t154 model. It would be interesting to
compare the two models by quantifying this difference via the type
of directional multi-point correlators devised by Doliwa and Heuer
\cite{doliwa1999origin}.
It is not clear if the difference between the models is related to
the fundamental distinction between LGMs and KCMs or just the specifics
of the particular models considered. In particular, the t154 is a
multi-component model, unlike the Kob-Andersen model. The string-like
motion on short time scales seems to occur predominantly on the rather
rough boundaries of slow clusters\cite{appignanesi2006democratic}.
This behavior, reminiscent
of the picture of dynamic heterogeneity that put forward by Stillinger
\cite{stillinger1988relaxation},
might be strongly influenced by compositional heterogeneity. A useful
way to address general issues related to how the initial configuration
constrains subsequent dynamics would be a systematic iso-configurational
ensemble analysis comparing LGMs and KCMs
\cite{widmercooper2004reproducible}. This will be the topic
of a future publication \cite{darst2009tbp}.

\begin{figure}
\centering
\subfigure[]{
  \includegraphics[width=\halfplotwidth,angle=-90]{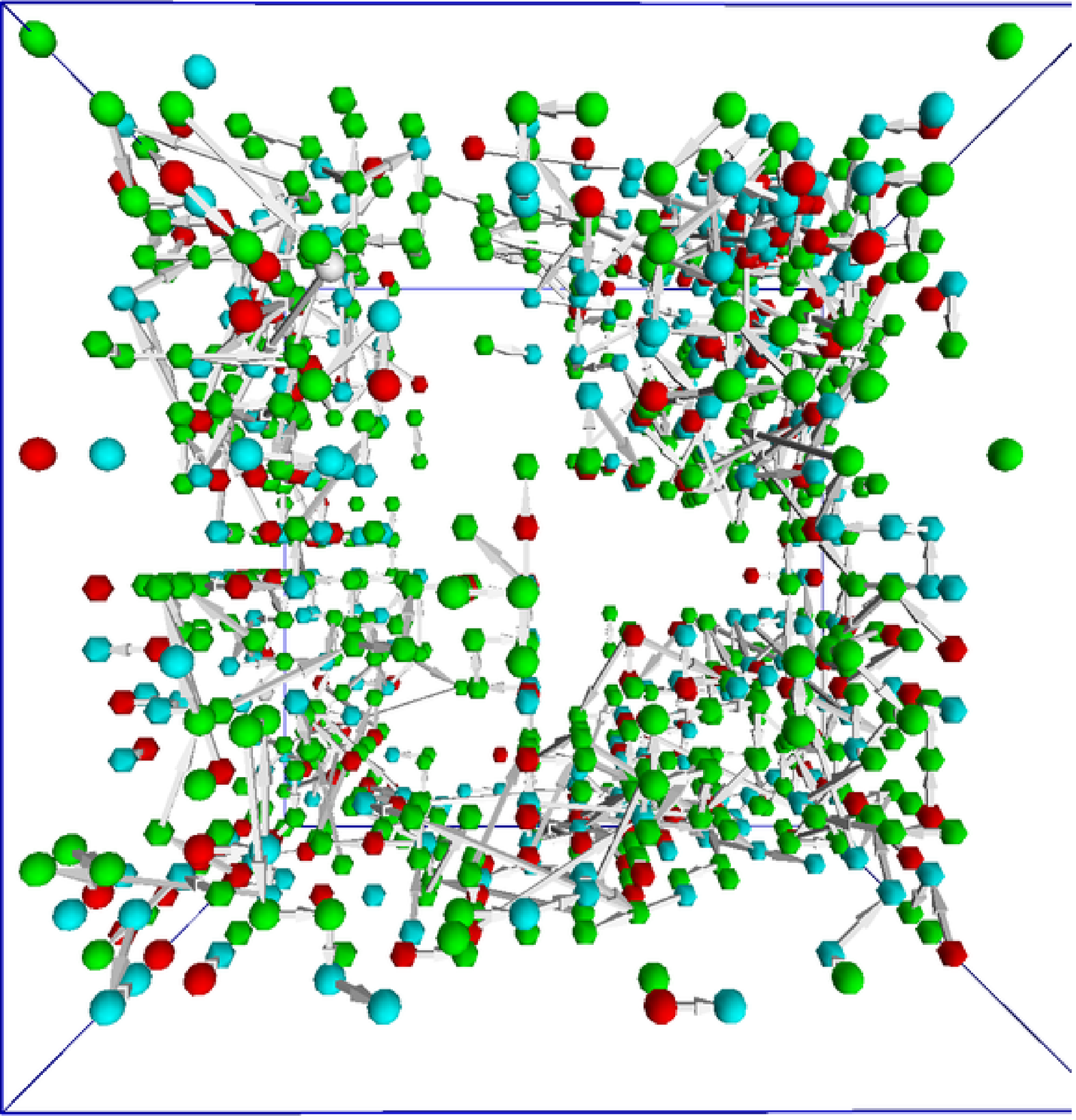}}
\subfigure[]{
  \includegraphics[width=\halfplotwidth,angle=-90]{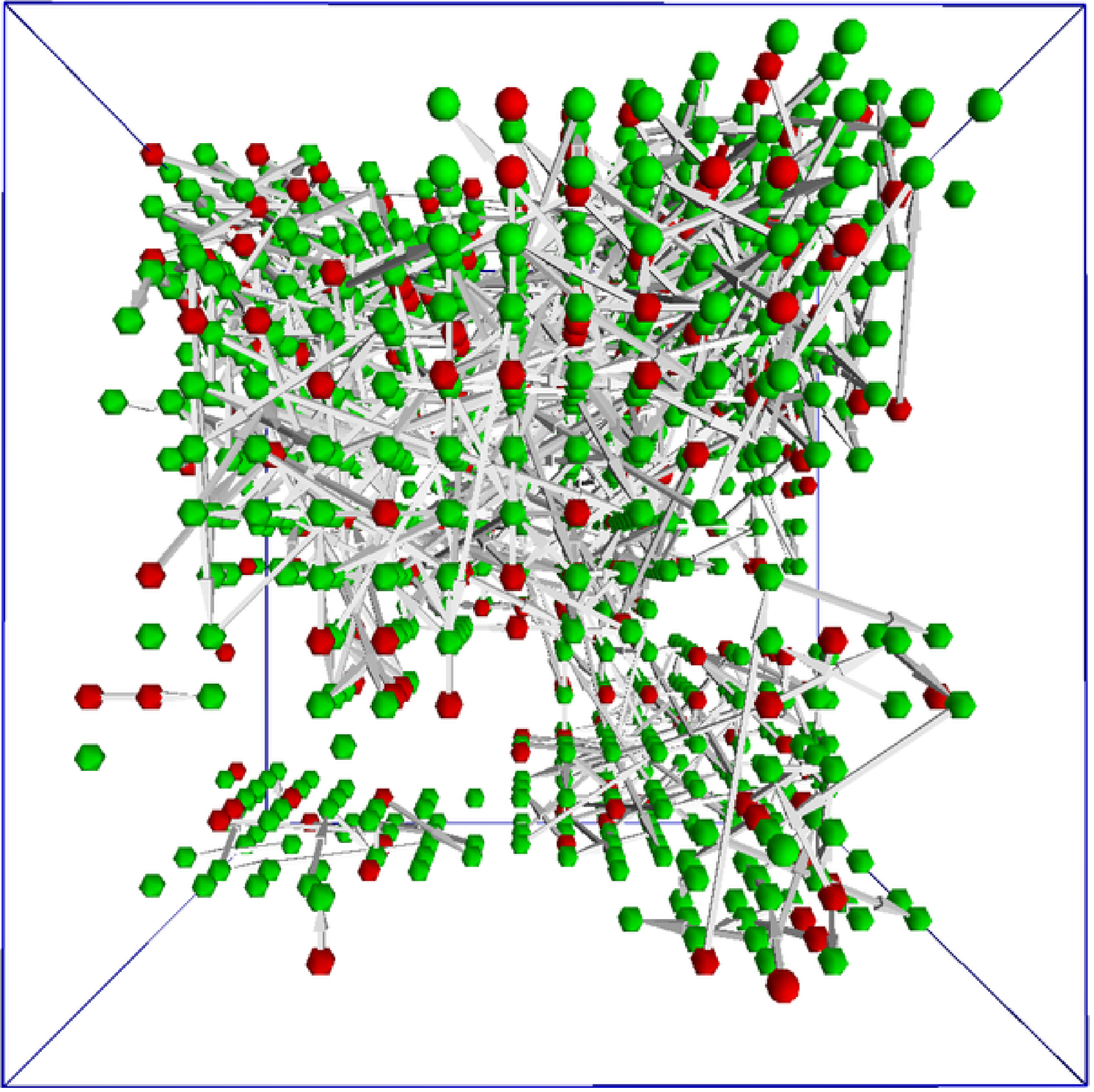}}
\caption{Examples cluster shapes in the \textbf{(a)} the t154, model,
  density $\rho=.5400$ and
  \textbf{(b)}
  the Kob-Andersen model, density $\rho=.8500$. Arrows indicate motion
  between initial and final times. Time separation is 1/10th of the
  $\alpha$-relaxation time. In the t154 model, we see more fractal and
  disconnected clusters, while in the KA model, mobile domains tend to be
  smoother clusters.}
\label{fig:clusters}
\end{figure}

In the next few sections, we discuss how some of the most important indicators
of dynamical heterogeneity in supercooled liquids manifest in the
t154 model. The quantities that we discuss are the magnitude of violations
of the Stokes-Einstein relation, exponential tails (indicative of
hopping transport) in the van Hove function, the existence of a Fickian
length scale and the development of a dynamical length scale quantified
by the multi-point function $S_4(q,t)$.
Unless otherwise stated, specific correlation functions and transport
coefficients are calculated with respect to type-2 particles.

\subsection{Stokes-Einstein Violation}

In typical fluids a mean-field linear-response relationship asserts
that the product of the tracer particle diffusion constant and the
fluid viscosity divided by the temperature is a constant
\cite{balucani1994dynamics}. This connection
between diffusion and dissipation is known as the Stokes-Einstein
relationship, and empirically is known to hold even at the atomic
scale in liquids over a wide range of densities and temperatures.
In supercooled liquids, the Stokes-Einstein relation generally does
not hold \cite{chang1997heterogeneity, cicerone1996enhanced,
  jung2004excitation,
  xia2001diffusion, tarjus1995breakdown, berthier2005length,
  stillinger1994translation, liu1996enhanced}. In fact,
the product of the diffusion constant and the viscosity
of a liquid may exceed that expected from the Stokes-Einstein relation
by several orders of magnitude close to the glass transition. There
are many theoretical explanations for Stokes-Einstein violations in
supercooled liquids, which essentially all invoke dynamical heterogeneity
as the fundamental factor leading to the breakdown of the simple relationship
between diffusion and viscosity. It should be noted that similar relationships
hold between the diffusion constant and the self and collective time
constants associated with the decay of density fluctuations. In this
work we focus on the relaxation time of the self-intermediate scattering
function defined above as our proxy for the fluid viscosity.

\begin{figure}
\centering \includegraphics[angle=-90,width=\plotwidth]{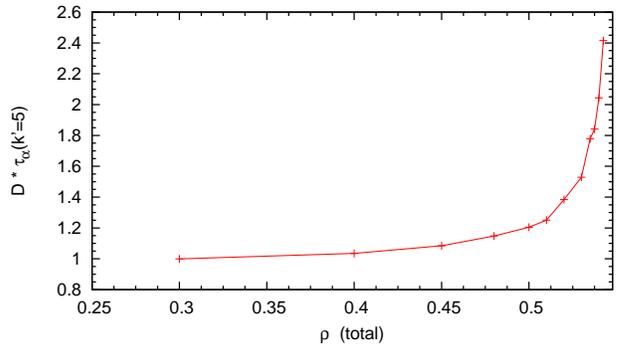}
\caption{Violation of the Stokes-Einstein relation, $D\tau_\alpha\sim
  \mathbf{constant}$, using $\tau_{\alpha}$ at $k'$=5.
Data has been normalized to $D{\tau_\alpha}=1$ at the lowest density.}
\label{fig:stokeseinstein}
\end{figure}

It is well known that the product $D \tau_{\alpha}$, where
$\tau_{\alpha}$ is the $\alpha$-relaxation time of the self-intermediate
scattering function shows a strong temperature/density dependence
in both realistic atomic simulations as well as in the class of KCMs
that describe fragile glass-forming liquids. No direct studies of
this quantity have been made in LGMs. The LGM of Coniglio and coworkers
would appear to show essentially no Stokes-Einstein violations because
the diffusion constant and the relaxation time may both be fit to
power laws with exponents that have, within numerical accuracy, the
same magnitude \cite{ciamarra2003lattice, pica2003monodisperse}. This,
however is not surprising since many of the features
of the model resemble those of a strong glass-forming system, where
violations of the Stokes-Einstein relation are, at most, weak. The
features of the t154 model with regard to non-exponential relaxation
and the density dependence of the relaxation time $\tau_{\alpha}$ indicate
that this model behaves more like a fragile glass former. Thus, we
expect clear violations of the Stokes-Einstein relation. Indeed, as
shown in Fig. \ref{fig:stokeseinstein}, $D\tau_{\alpha}$ increases
markedly as density is increased. Over the range densities that we
can access, the magnitude of the violation is very similar to
that seen in the canonical Kob-Andersen Lennard Jones mixture over
a comparable range of changes in relaxation time
\cite{chaudhuri2007universal}. Interestingly, violations
begin to become pronounced at densities similar to where the relaxation
times and stretching exponents become strongly sensitive to increased
density. Thus, a consistent onset density is observed as in more
realistic atomistic systems.

\subsection{van Hove Function}
It is now rather well established that an additional
``quasi-universal'' feature of dynamical heterogeneity near the glass
transition is contained in the shape of the real-space van Hove
function \cite{stariolo2006fickian, chaudhuri2007universal,
  saltzman2008large, swallen2009self}.  In particular it has been
argued the tails of the self van
Hove function should be approximately exponential in form.  These ``fat
tails'' imply that the rare particles that do undergo large
displacements exist in populations in excess of what would be expected
in a purely Gaussian displacement distribution. While non-Gaussian
tails should be expected of any distribution for the wings that fall
outside of limits of bounds set by the Central Limit Theorem, the
palpable exponential tails in supercooled liquids imply large
non-Gaussian effects indicative of transport that is strongly effected
by heterogeneous hopping motion.

\begin{figure}
\centering \includegraphics[angle=-90,width=\plotwidth]{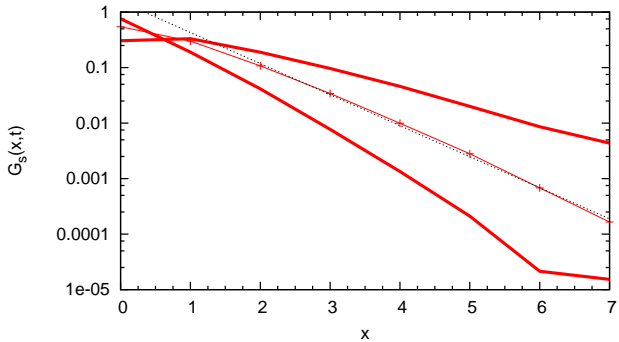}
\caption{van Hove function for $\rho=.5375$ and various times. Distances are
measured independently along each coordinate axis.
The times plotted, from left to right, are $10^{5}$, 316227
(approx. the $\alpha$-relaxation time), and $10^{6}$.  An exponential
fit to the tail of the $t=316227$ case is shown by a dotted line.}
\label{fig:vanhove}
\end{figure}

Here, we demonstrate that such effects occur in the t154 model in a
manner similar to that seen both in experiments in colloidal and
granular systems as well as in computer simulations of atomic systems.
Fig. \ref{fig:vanhove} shows the self part of the real-space van Hove
function,
\begin{equation}
  \label{eq:vanHove}
  G_s(x,t) = \left<
    \delta\left(x-\left|\mathbf{\hat{x}}\cdot\left(
       \mathbf{r}_i(t) - \mathbf{r}_i(0)
     \right)  \right| \right)
  \right>,
\end{equation}
for the type two particles in the t154 model.  Because we are on a
lattice, we restrict our distances along the three coordinate axes
$\mathbf{\hat{x}}$ individually in our calculation.  We see that for
times of the order of the $\alpha$-relaxation time, these tails are
clearly visible.  For very long or short time scales, the shape of the
tail deviates somewhat from the more exponential form exhibited at
intermediate times.  This behavior is quite similar to that seen in
simulations of atomistic systems \cite{chaudhuri2007universal,
  saltzman2008large}, and is fully consistent with the
behavior found in KCMs \cite{berthier2005length}.

\subsection{Fickian Length}
\begin{figure}
\centering
\includegraphics[angle=-90,width=\plotwidth]{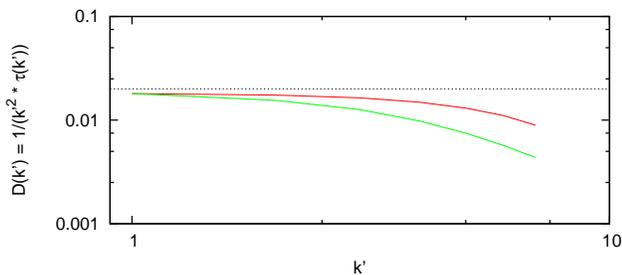}
\caption{$k$-dependent diffusion constant
$=1/\left(k'^{2}\tau_{\alpha}(k)\right)$.
Densities of .3000 (upper) and .5425 (lower).  The higher density
curve is multiplied by a scale factor of $2.992\times10^{5}$
for ease of comparison.  A dotted flat line is included for reference
of behavior expected in the purely Fickian case.}
\label{fig:diffuse_k}
\end{figure}

Related to the existence of excess tails in the van Hove function is
the existence of a length scale that characterizes the anomalous
transport.  More specifically, the exponential tails in the van Hove
function are distinguished from the Gaussian form of the displacement
distribution obtained at relatively short distances for fixed
times. The crossover from Fickian to non-Fickian behavior should be
characterized by time scales as well as {\em length scales} over which
this crossover occurs.  A non-Fickian length scale may be defined by
examining the $k$-dependent diffusion constant
$D(k)=\frac{1}{\tau_{\alpha}k^{2}}$ \cite{berthier2004time,
  pan2005heterogeneity, szamel2006time}. The wavevector that characterizes
the crossover from the expected diffusive behavior to an anomalous
regime is inversely related to such a length scale.  In
Fig. \ref{fig:diffuse_k} we plot $D(k')$.  Clearly, as the density is
increased, the length scale separating the Fickian and non-Fickian
regimes increases.  This behavior is consistent with that found in
KCMs and simulations of atomistic glass-forming liquids.  It should be
noted that Stokes-Einstein violations, the development of exponential
tails in the self van Hove function, and a well-developed Fickian
length scale are all manifestations of related aspects of dynamically
heterogeneous motion in supercooled liquids \cite{chaudhuri2007universal}.

\subsection{$\chi_{4}$ and $S_{4}$ Fluctuation Measures}
The Fickian length scale is merely one length scale that arises
naturally in systems where dynamics become increasingly heterogeneous.
Perhaps more fundamental is the growth of dynamical length scales
associated with multi-point correlations of the dynamics. Supercooled
liquids do not show simple static correlations that would indicate a
growing correlation length.  It should be noted that this does not
exclude growing static correlations of a more complex kind, for
example point-to-set correlations \cite{bouchaud2004adam,
  mezard2006reconstruction,
  biroli2008thermodynamic}.  Regardless, cooperativity in
dynamics may be measured via first defining a local overlap function
\cite{garrahan2002geometrical, yamamoto1998dynamics,
  lacevic2003spatially, franz2000non, berthier2007spontaneous}
\begin{equation}
  \delta f_k(q,t) = \frac{1}{N}
     \sum_i e^{i \mathbf{q}\cdot\mathbf{r}_i(0)}
       \left[
         \cos\left(\mathbf{k}\cdot\left(\Delta\mathbf{r}_i(t)\right)\right)
         - F_s(k,t)
       \right]
\end{equation}
where $\Delta\mathbf{r}_i(t) = \mathbf{r}_i(0)-\mathbf{r}_i(t)$.
$f_k(q,t)$ is defined for one configuration, and the average is over
all $\mathbf{k}$ and $\mathbf{q}$ with the
magnitudes $k$ and $q$.  Then, $S_4(q)$ is defined as
\begin{equation}
  S_4(q) = N \left< \left|\delta f_k(q,t)\right|^2 \right>
\end{equation}
where this average is over the most general ensemble of configurations
\cite{berthier2007spontaneous}.
The $\chi_4$ value is defined as the limit $S_4(q\to0)$.
$\chi_4(t)$ may be calculated strictly at $q=0$ from
\begin{equation}
  \chi_4(t) = N \left< \left|\delta f_k(q=0,t)\right|^2 \right>
\end{equation}
where the average is over the entire ensemble and all $\mathbf{k}$
consistent with the magnitude of k.  Note that, as discussed in
\cite{berthier2007spontaneous}, the value of $\chi_4(t)$ computed in this
manner is a lower bound for the extrapolation of $S_4(q\to0,t)$.

The quantity $S_{4}(q,t)$ is a multi-point dynamical
analog of $S(q)$.  Just as the low $q$ behavior of $S(q)$ indicates a
growing (static) length scale in systems approaching a second order
phase transition, scattering from dynamically heterogeneous regions
undergoing cooperative motion will manifest growth in the amplitude of
the low $q$ region of $S_{4}^{ol}(q,t)$, indicative of the
size scale of the dynamical correlations for systems approaching the glass
transition.

\begin{figure}
\centering
\includegraphics[angle=-90,width=\plotwidth]{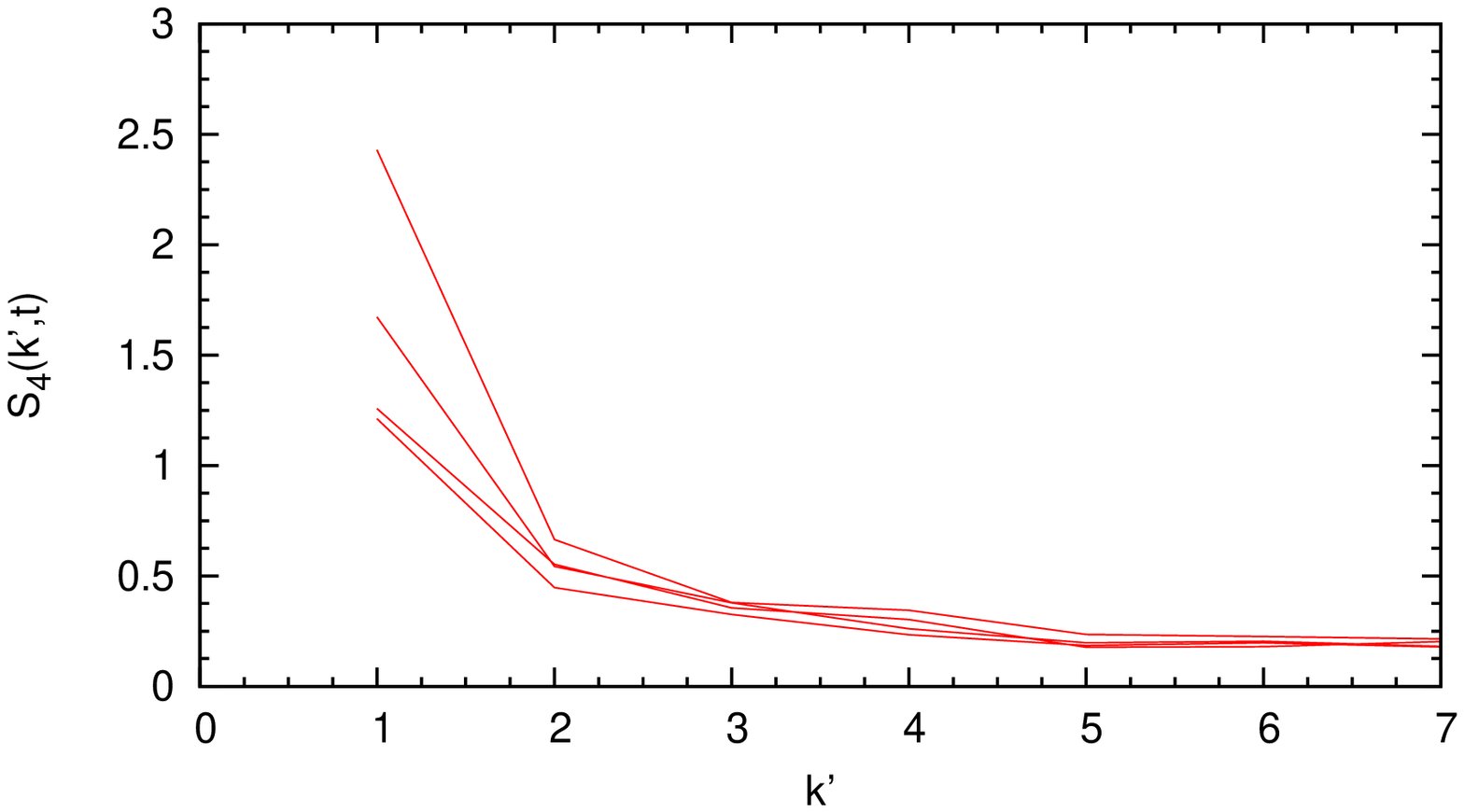} \tabularnewline
\includegraphics[angle=-90,width=\plotwidth]{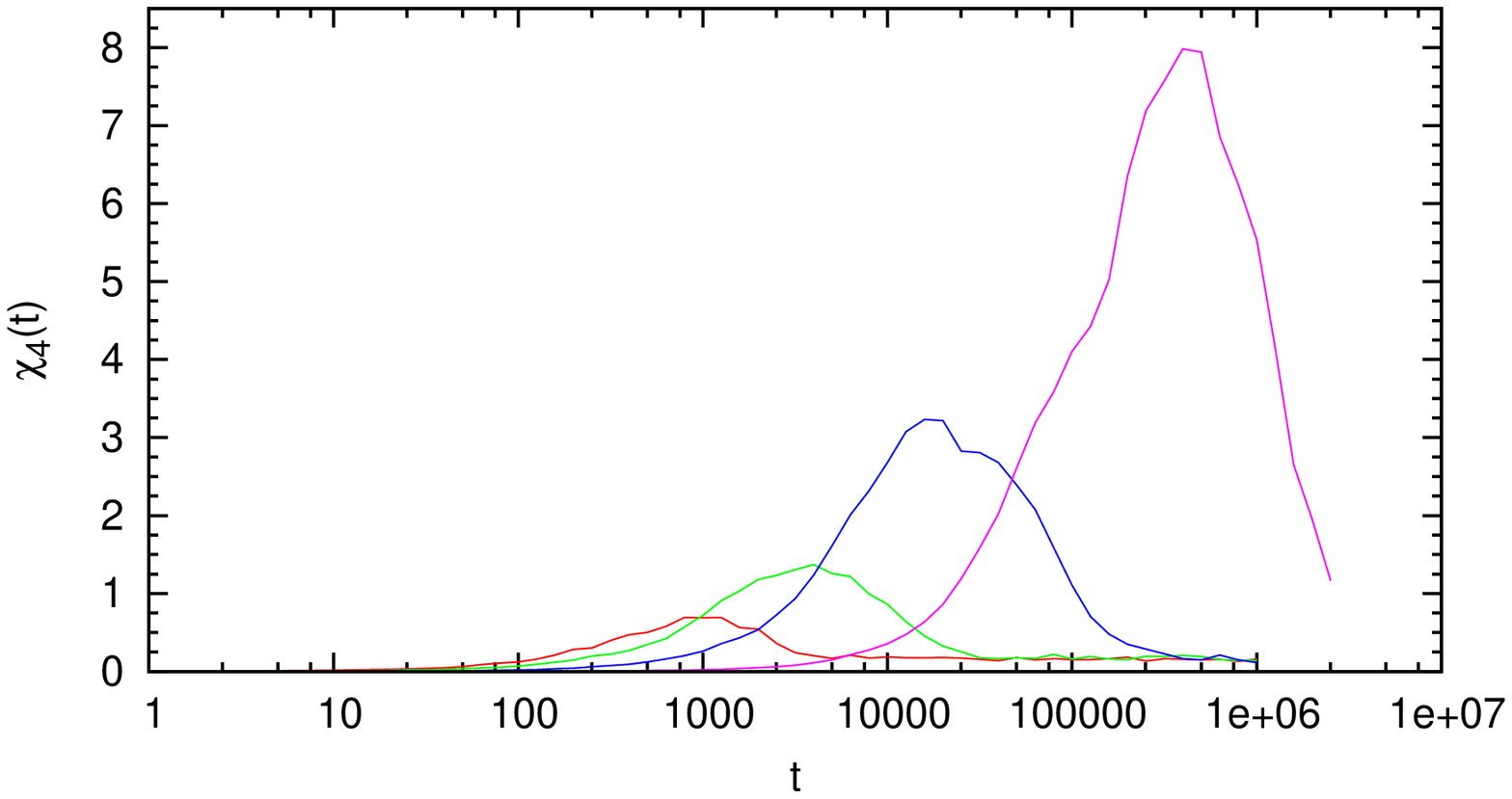} \tabularnewline
\caption{
  \textbf{Top:}
  Top:Plot of $S_4(q,t)$ at $\tau_\alpha$ for densities 0.51, 0.52,
  0.53 and 0.54.
  \textbf{Bottom:} Plot of $\chi_4(t)$ for the same densities. Peak
  values correspond to lower bounds for of the value of $S_4(q,t)$ in
  the upper panel at $q=0$.
}
\label{fig:chi4}
\end{figure}

The behavior of the quantity $S_{4}^{ol}(q,t)$ is shown in
Fig. \ref{fig:chi4}. Only type-2 particles have been used in the
calculation.  As can clearly be seen, for densities above $\rho \sim
0.5$ which constitutes the onset density of this system, the low $q$
behavior shows a marked upturn as $q \rightarrow 0$.  The growth of
$S_{4}^{ol}(q,t)$ as $q \rightarrow 0$ as density is
increased suggests a growing length scale as supercooling
progresses. This non-trivial behavior is what is found in atomistic
simulated systems.  Future work will be devoted to a precise
characterization of the length scale that may be extracted from
$S_{4}^{ol}(q,t)$ in the t154 model so that a comparison may
be made with recent work detailing the behavior of this length in
realistic off-lattice systems \cite{stein2008scaling, karmakar2009growing}.

\section{Conclusion}
\label{sec:conclusion}

In this paper we have presented a new LGM based on the original Biroli-M\'ezard
model \cite{biroli2001lattice}.
Via the introduction of an additional species of particle,
we have demonstrated that our model is stable against crystallization.
This fact allows us to study sufficiently high density configurations
that manifest features of dynamical heterogeneity. Unlike some previous
LGMs, our model exhibits the canonical features of a fragile glass-former.
In terms of the gross features of relaxation behavior, our LGM shows
behavior similar to the standard Kob-Andersen Lennard-Jones (KALJ)
mixture. In particular, we find that the degree of violation of the
Stokes-Einstein relation and the magnitude of stretching in the decay
of the self-intermediate scattering function track the relaxation
times at densities above the onset of supercooling in a manner
consistent with that seen in the KALJ system. Features of dynamical
heterogeneity such as exponential tails in the van Hove function, the growth
of a dynamical length scale as quantified by the function $S_4(q,t)$,
Stokes-Einstein violations and the emergence of a Fickian length scale
all occur in a manner expected from experiments and simulations of
fragile glass-forming liquids.

The similarity between the description of dynamic heterogeneity found
in KCMs and LGMs stands in stark contrast to the underlying foundations
of the models themselves. As emphasized in the introduction, KCMs
are based on a constrained dynamics for which the number of available
dynamical paths leading to relaxation becomes increasingly rare as
the density increases and the number of defects decrease. In KCMs
all real-space configurations at a fixed number of defects (excluding
rare blocked configurations) are equally likely. On the
other hand LGMs are based on transitions between real-space configurations
that become increasingly scarce as the density is increased.  This is
not to say that there is not a facilitated-like dynamics in LGMs.
On the contrary, as we have demonstrated in sec. \ref{sec:dynamics},
local and sometimes
anisotropic dynamics may be generated naturally in LGMs without the
explicit introduction of facilitating defects. An important message
that emerges from this study is that {\em the phenomenology
of dynamic heterogeneity is not sufficient to distinguish pictures
or validate models
based on transitions between sets of states in configuration space
from those based on sets of paths in space-time}.

How then might these pictures be differentiated? While contrasting
competing models that generate seemingly similar dynamical behavior
is a difficult endeavor, several possible studies might be useful
for this task. Here we outline four avenues that could provide key
information that distinguish the purely dynamical picture from one
based on transitions thermodynamic states.

a) \emph{The mosaic length scale:} The Random First Order Theory (RFOT) of
Wolynes and coworkers posits the existence of a static length scale
which is defined by the region over which particles are pinned by
the surrounding self-generated amorphous configuration
\cite{lubchenko2007theory, bouchaud2004adam, biroli2008thermodynamic}.
This length
scale also exists in KCMs, but it is decoupled from the relaxation
dynamics of the system \cite{jack2005caging}.
Recent atomistic computer simulations have
successfully located the mosaic length scale \cite{biroli2008thermodynamic}.
It would be quite useful
to perform an analysis similar to that devised by Jack and Garrahan
for LGMs \cite{jack2005caging}.
Since LGMs are based on the entropy of real-space configurations,
it is expected that here the mosaic length does couple to the glassy
dynamics. Since LGMs are much simpler than atomistic off-lattice
models, the direct study of the mosaic length (and point-to-set correlations
in general) in LGMs might provide key avenues for the testing of the
putative coupling between relaxation and such length scales in simulated
atomistic systems.

b) \emph{Correlations between configurational entropy and dynamics:} Empirical
correlations between the configurational entropy and the $\alpha$-relaxation
time have been noted for many years, and this correlation
lies at the heart of several prominent theories. Such correlations are
still widely debated, but seem to hold at least crudely in many
glass-forming systems \cite{richert1998dynamics, martinez2001thermodynamic}.
LGMs should be expected to exhibit such
correlations, while it is known that KCMs do not exhibit such
correlations. Recently Karmakar {\em et al.} purported to show that
finite-size effects of the $\alpha$ relaxation time follow precisely the
Adam-Gibbs relation between the configurational entropy and the
$\alpha$-relaxation time in the KALJ system \cite{karmakar2009growing}.
If true, such correlations would
be a challenge to KCMs, since it is difficult to envision how the
configurational entropy would track the $\alpha$-relaxation time for
different system sizes if it were not a crucial component of
relaxation phenomena.  Such correlations, however, are subtle to
measure since the Adam-Gibbs relationship is an exponential one and
the apparent correlation could depend on the somewhat indirect
computational method used in \cite{karmakar2009growing} to define
the configurational
entropy. It would be most useful to investigate such effects in the
simpler LGMs, which might provide a cleaner means of isolating the
configurational entropy. It should be noted that finite size effects
do appear to follow an approximate Adam-Gibbs relationship in at least
one other lattice model \cite{crisanti2000potential}.
Such studies might spur more detailed
investigations in simulated atomistic systems thus allowing for a
clear comparison between LGMs, KCMs and more realistic systems.

c) \emph{Single-particle and collective predictability ratios:} In an important
piece of work, Jack and Berthier devised metrics that access the degree
to which single particle and collective dynamics are deterministically
predicted by a set initial configuration over a given time scale
\cite{berthier2007structure}.
KCMs and LGMs differ in how allowed configurations are constructed.
KCMs have explicit defects, while configurations in LGMs are determined
by global constraints, and thus do not contain explicit defects. Since
the very composition of initial conditions differ markedly in these
models, one expects that the metrics defined by Jack and Berthier
would behave differently in KCMs and LGMs. Thus, it would be very
profitable to examine the density and temperature dependence of the
single particle and collective predictability ratios in KCMs and LGMs
as a possible means of distinguishing between state-based, and dynamical
constraint-based pictures \cite{darst2009tbp}.

d) \emph{Evolution of the facilitation mechanism approaching the glass
transition:} Although in both KCM and LGM pictures facilitation plays
an important
role in the relaxation of the system, a peculiar and different
temperature and density evolution is expected. In particular, in the
KCM picture, facilitation is due to the motion of mobility regions or
defects.  Dynamics slows down, and concomitantly dynamic heterogeneity
increases, because these regions become rarer approaching the glass
transition.  A crucial assumption is that these defects are conserved
or at least that non-conservation is a rare event that becomes rarer
at lower temperature/high density.  These assumptions impose important
constraints on the evolution of the facilitation mechanism.
Thus, it would be very
interesting to examine this issue for example using the cluster
analysis developed in \cite{candelier2009building} to study the
relaxation dynamics of granular systems.

Investigation of these and other studies aimed at distinguishing the
underlying pictures that LGMs and KCMs are based on will be the subject
of future work.

\begin{acknowledgements}
  RKD would like to thank the John and Fannie Hertz Foundation for
  research support via a Hertz Foundation Graduate Fellowship.  RKD and
  DRR would like to thank the NSF for financial support.  We would
  like to thank Ludovic Berthier, Joel Eaves, Peter Harrowell, Robert
  Jack, Peter Mayer and Marco Tarzia for useful discussions.
\end{acknowledgements}

\bibliography{references,footnotes}{}
\bibliographystyle{apsrev}

\end{document}